# Concept study for a high-efficiency nanowire-based thermoelectric

M. F. O'Dwyer[1], T. E. Humphrey[2] and H. Linke[3]

[1] School of Engineering Physics and Institute for Superconducting and Electronic Materials, University of Wollongong, Wollongong NSW 2522, Australia
[2] Département de Physique Théorique, University of Geneva, CH – 1211, Geneva, Switzerland
[3] Materials Science Institute and Physics Department, University of Oregon, Eugene, OR 97403-1274, U.S.A.

**Abstract**

Materials capable of highly efficient, direct thermal-to-electric energy conversion would have substantial economic potential. Theory predicts that thermoelectric efficiencies approaching the Carnot limit can be achieved at low temperatures in one-dimensional conductors that contain an energy filter such as a double-barrier resonant tunneling structure. The recent advances in growth techniques suggest that such devices can now be realized in heterostructured, semiconductor nanowires. Here we propose specific structural parameters for InAs/InP nanowires that may allow the experimental observation of near-Carnot efficient thermoelectric energy conversion in a single nanowire at low temperature.

## 1. Introduction

Thermoelectric devices can be used as power generators or refrigerators and have many advantages over other energy converters, including high reliability, long life, lack of moving parts, low maintenance requirements, scalability, possible miniaturization and absence of emissions [1,2]. Whereas the relatively low efficiency of current practical thermoelectric devices limits their use to niche applications, a breakthrough in thermoelectric technology would have substantial economic impact. Recent theory predicts that the use of nanowires in thermoelectric devices could have significant advantages. Specifically, it has been predicted that low-dimensional conductors in which mobile electrons are restricted to a very narrow energy range can achieve electronic energy-conversion efficiencies approaching the Carnot limit [3,4].

The recent advances in nanowire growth techniques now give access to high-mobility, heterostructured nanowires with a tuneable density of states [5-9]. Here, we develop specific heterostructure parameters for an InAs/InP nanowire heterostructure that should be suited to testing the above theoretical predictions. We consider a wire with a single embedded quantum dot, in which electrons tunnel via a single resonant level between hot and cold electron reservoirs. We propose an experiment that will be performed at temperatures below 10 K where heat transfer between electrons and phonons is suppressed, allowing efficiency gains due to optimization of the electronic structure to be probed in the absence of phonon-mediated heat transport. The objective of the planned experiment is the proof-of-principle demonstration that energy-filtering in 1D conductors can significantly increase the efficiency of thermoelectric energy conversion.

## 2. Thermoelectrics overview

The thermoelectric figure of merit, used to characterize the performance of a thermoelectric device, is given by

$$ZT = \frac{\alpha^2 \sigma}{\kappa_e + \kappa_l} T \qquad (1)$$

where $\alpha$ is the Seebeck coefficient, $\sigma$ is the electrical conductivity, $\kappa_e$ is the electronic thermal conductivity, $\kappa_l$ is the lattice thermal conductivity and $T$ the temperature. It is related to efficiency via [10]

$$\eta = \frac{P_{out}}{\dot{Q}_H} = \eta_C \frac{\left[\sqrt{1+Z\overline{T}} - 1\right]}{\sqrt{1+Z\overline{T}} + T_C/T_H} \qquad (2)$$

where $\eta_C = 1 - T_C/T_H$ is the Carnot efficiency, $T_{H/C}$ are the hot/cold reservoir temperatures with average $\overline{T}$, $P_{out} = IV_{bias}$ is the power output of the device and $\dot{Q}_H$ is the heat flow from the hot reservoir. A good thermoelectric device therefore has high Seebeck coefficient (thermopower) and electrical conductivity, and low thermal conductivity. Energy conversion with efficiency near the Carnot limit corresponds to a diverging $ZT$.

The figure of merit for bulk semiconductor based devices has been relatively stagnant over the last forty years, with $ZT \sim 1$ and efficiencies of about 10% of the Carnot limit in commercial devices being achieved with optimized Bismuth Telluride based alloys [11]. Recently however, the possibility of tailored electrical and thermal properties offered by nanostructured materials has attracted considerable attention. Hicks and Dresselhaus proposed that the reduced dimensionality of superlattices could be used to enhance the electronic density of states in energy ranges advantageous for thermoelectric power generation [12], and various authors have shown that thermal conductivity in superlattices is much lower than pure bulk materials and can be lower than the alloy limit [13-15]. Harman et al. have reported a $ZT$ of up to 1.6 using a PbSeTe based quantum dot superlattice [16]. The highest reported room-temperature $ZT$ to date, ~ 2.4 by Venkatasubramanian et al., was achieved using a $p$-type $Bi_2Te_3/Sb_2Te_3$ superlattice and was attributed to good electrical transport and very low thermal conductivity [17].

Nanowires offer the possibility of yet further improvements in the figure of merit. Hicks and Dresselhaus predict $ZT \approx 6$ for $Bi_2Te_3$ based nanowires compared to 2.5 and 0.5 for superlattice and bulk systems respectively [18]. This significant improvement was attributed primarily to changes in the density of states in addition to reduced lattice thermal conductivity due to increased phonon surface scattering. Lin and Dresselhaus presented a model which described thermoelectric transport in superlattice nanowires, again predicting that significantly higher figures of merit could be achieved in such devices [19].

Thermal-to-electric energy conversion with an efficiency near the Carnot limit would require negligible lattice heat conduction, and reversible electron transport between the hot and cold reservoirs, that is, current flow without entropy production. We recently showed that the latter conditions can indeed be achieved in principle if the only electrons that can travel between the heat reservoirs are those that have a specific energy given by [3]

$$E_0 = \frac{\mu_C T_H - \mu_H T_C}{T_H - T_C} \quad (3)$$

where $\mu_{H/C}$ are the hot/cold reservoir chemical potentials. Electrons at this energy are said to be in 'energy-specific equilibrium' [4] because they can move reversibly between the two reservoirs. If electrons are transmitted at an energy infinitesimally close to $E_0$, the efficiency of energy conversion will approach the Carnot value.

To see how the above condition can be implemented, consider the system shown in figures 1 and 2. A hot and a cold electron reservoir, each populated with electrons with energy distributions given by Fermi-Dirac statistics (see appendix, equation A.3), are connected by a double-barrier resonant tunneling structure that restricts electron flow to an energy range of width $\delta E$, around a specific energy $E_{res}$. Importantly, the length ($w + 2b$) of this "energy filter" is chosen to be much shorter than the electron mean free path for inelastic scattering, and an electron leaving one reservoir at $E = E_{res}$ will arrive in the other reservoir at the same energy, before equilibriating in the reservoir. The net flow of electrons will be from the reservoir with the higher occupation of states at $E_{res}$. When the two electron reservoirs have different temperatures and electrochemical potentials, $E_0$ is the energy at which the occupation of states cross, as shown in figure 2. Net current will flow against the bias voltage from hot to cold when $E_{res} > E_0$ (power generation), and from cold to hot when $E_{res} < E_0$ (refrigeration), as indicated in figures 2a and b. In an experiment, the relative positions of $E_0$ and $E_{res}$ can be changed through varying $E_0$ by changing $\mu_C$ and $\mu_H$ using a bias voltage or a gate voltage. In an ideal device, the highest power generation and refrigeration efficiencies will be obtained when $E_{res}$ is just higher than or just lower than $E_0$, respectively.

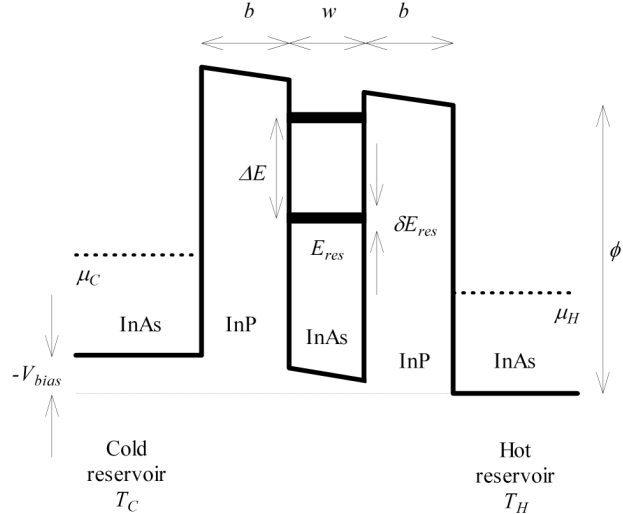

**Figure 1.** Band diagram showing the double barrier structure and key system parameters used in the model. $\mu_{H/C}$ are the chemical potentials of the hot/cold reservoirs, $T_{H/C}$ are the temperatures of the hot/cold reservoirs, $E_{res}$ is the resonant transmission energy and has energy width $\delta E_{res}$, $\Delta E$ is the energy gap between the first and second resonance (if any), $\varphi$ is the conduction band offset, $V_{bias}$ is the bias voltage on the system applied between the source and drain, and $b$ and $w$ are the barrier and well widths, respectively.

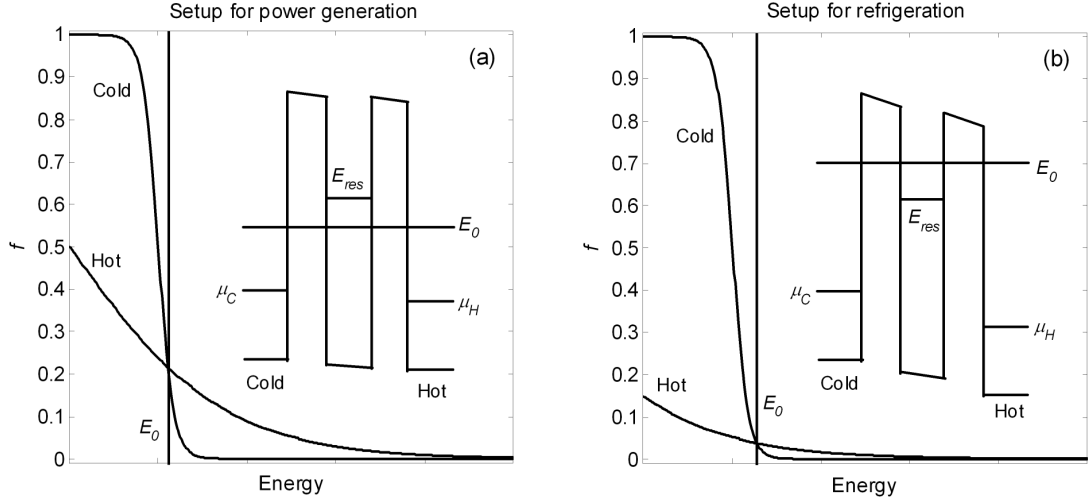

**Figure 2.** (a) The cold and hot reservoir Fermi-Dirac distribution functions at low negative bias, where $E_0 < E_{res}$, and the system acts as a power generator, with the heat gradient driving a current against the bias voltage from hot to cold. (b) At larger negative bias $E_0$ shifts to higher energy, sweeping through the resonance and the cold reservoir distribution function at $E_{res}$ is now larger than that of the hot reservoir causing current flow from cold to hot and the device to act as a refrigerator. Efficiency in each case is higher when $E_{res}$ is closer to $E_0$.

## 3. Why nanowires? Practical requirements for near-Carnot efficiency.

There are a number of requirements our system must meet in order to achieve an efficiency approaching the Carnot value.

First, the filtering of electron energies for transmission must occur according to their *total* energy. This is not possible in conventional 3D superlattices which filter electrons according to their energy in the direction of transport only, allowing energies in the other spatial dimensions to take on random values [20]. In a nanowire, electron energies are quantized in the direction perpendicular to transport. Should these energy levels be sufficiently spaced relative to the thermal energy, $k_BT$ (~1 meV at 10 K), transport in the nanowire attached to the double barrier will be restricted to a single level. Thus, the nanowire should be sufficiently thin that the first energy level is a few $k_BT$ from the second and will therefore dominate transport. If we have an InAs well with effective mass of $m^* = 0.023m_0$, the energy levels given by an infinite square well of width 40 nm [21] are approximately $E_1 = 10$ meV, $E_2 = 42$ meV and $E_3 = 94$ meV. These energy levels may be calculated more accurately by obtaining a solution to the Schrödinger equation which takes into account the cylindrical geometry of the nanowire [22], and by taking into account charging energies (Coulomb blockade) and electronic shell structure [23-25]. However, with tunable electrochemical potentials, the precise energy offset of that level is unimportant for the discussion below.

Second, the double barrier needs to be designed such that $\Delta E$ (as shown on figure 1) is at least a few $k_BT$ so that the first resonance dominates current flow. Furthermore, efficiency is expected to be higher for smaller $\delta E_{res}$ [26], which can be tuned by adjusting the barrier width $b$, as described below.

Third, heat leaked by the lattice must be very low if high efficiency is to be achieved. This is particularly important when $\delta E_{res}$ is small because then the electron current and the power of the device will be very low. In order to achieve ideal conditions for the proposed proof-of-principle experiment we propose to operate at temperatures below 10 K where, electron-phonon coupling in semiconductors becomes weak [27]. Under these conditions we can assume that electron-electron interactions establish a quasi-equilibrium electron temperature which is independent of the phonon temperature. One electron reservoir can be heated using a heating current without phonon-mediated heat flow to the cold electron reservoir.

## 4. Device performance and discussion

In the following we develop a specific device structure and perform numerical predictions of its performance as a power generator, using heat from the hot reservoir to drive a current against an applied bias voltage. Transmission probabilities through the structure have been calculated by obtaining a numerical solution to the time-independent Schrödinger equation based on Airy functions [28,29]. Electrical and heat currents have been calculated using a Landauer equation based method which is detailed in the appendix. We assume that electron transmission through the device is dominated by a single energy level as demonstrated in references [6,25].

For the calculations to follow we use the geometry as shown in figure 1 with symmetric barriers, and $\varphi = 0.57$ eV [5], $m^*_{InAs} = 0.023 m_e$, $m^*_{InP} = 0.08 m_e$, $T_H = 10$ K, $T_C = 1$ K and electrochemical potentials are tuned for maximum performance unless stated otherwise.

The sign of the bias voltage is defined in figure 1, and positive current is from right to left. Under conditions where the device acts as a power generator, negative bias is used. The heat driven flow of negatively charged electrons from the hot reservoir on the right to the cold reservoir on the left is against the bias, that is, electric power is generated.

The width, $w$, of the well in the double barrier structure determines the number and energy of resonances. We find that for $w < 4$ nm the efficiency is low because $\delta E_{res}$ is too large and $E_{res}$ is too high, allowing over-barrier emission. Too large $w$ leads to many resonances, possibly so close together that more than one is actively involved in transport. Well widths between 5 nm and 9 nm give good results and we shall perform calculations with $w = 7$ nm unless otherwise noted. Figure 4(a) shows the transmission probabilities in this range.

As shown in figure 4(a), $\delta E_{res}$ decreases as $b$ increases, becoming less than $k_B T_C$ for $b > 7$ nm. If the barriers are not wide enough, the resonance becomes broad and efficiency drops (figure 4(b)). However, if the barriers become too thick, the resonance becomes so narrow that the current becomes extremely small, possibly smaller than the threshold of measurement. These phenomena are shown in figure 4(b), where we also notice that the predicted efficiency approaches the Carnot value when $b > 8$ nm. For the further calculations we choose $b = 7$ nm, which yields decent current generation (order of 10 pA) but still with efficiency quite close to the Carnot limit.

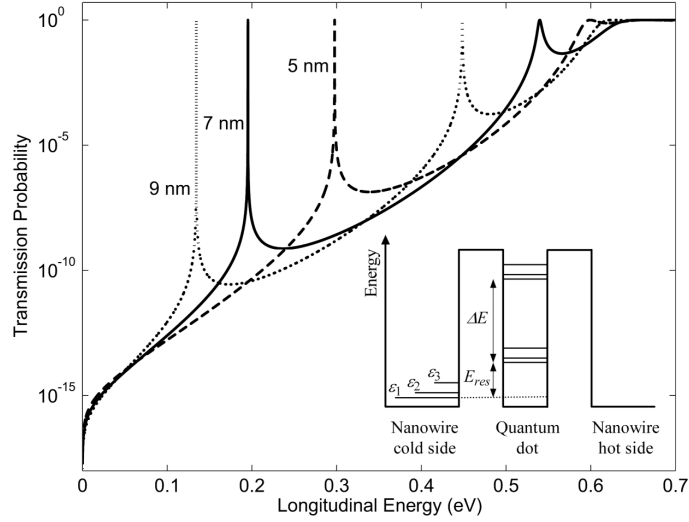

**Figure 3.** The transmission probabilities for double barrier structures with $b = 20$ nm and $w = 5$ nm, 7 nm and 9 nm. (Inset) Longitudinal energies, $E_{res}$ and $E_{res} + \Delta E$, are quantized due to the double barrier structure and transverse energies, $\varepsilon_{1,2,3}$, due to the dimensions of the nanowire (adapted from reference [6]).

Figure 5 shows that the efficiency is very sensitive to changes in the electrochemical potential. Since experimental variations in device parameters such as effective masses and barrier geometry can significantly alter the resonant transmission energy the electrochemical potential must be tunable, which will be achieved via a back-gate. In figure 6 we see that the current increases as the electrochemical potential approaches the resonant transmission energy before decreasing and reversing direction as the device shifts from the power generation to refrigeration regime. Figure 6 (inset) shows that current decreases as the hot reservoir temperature decreases. At $V_{bias} = 0$ net electron flow is from the hot to cold reservoir for all $T_H$ since there is no bias retarding flow out of the hot reservoir. As $V_{bias}$ becomes larger net current flows from cold to hot at low $T_H$ because the emitted current from the hot reservoir, which is impeded by the bias voltage, is lower than that out of the cold reservoir. As $T_H$ increases current emitted from the hot reservoir increases and the device begins to generate power as the current emitted from the hot reservoir becomes larger than that emitted from the cold reservoir. Figure 7 illustrates the I-P and I-V characteristics with maximum power being achieved in the system at a bias of approximately -1 mV.

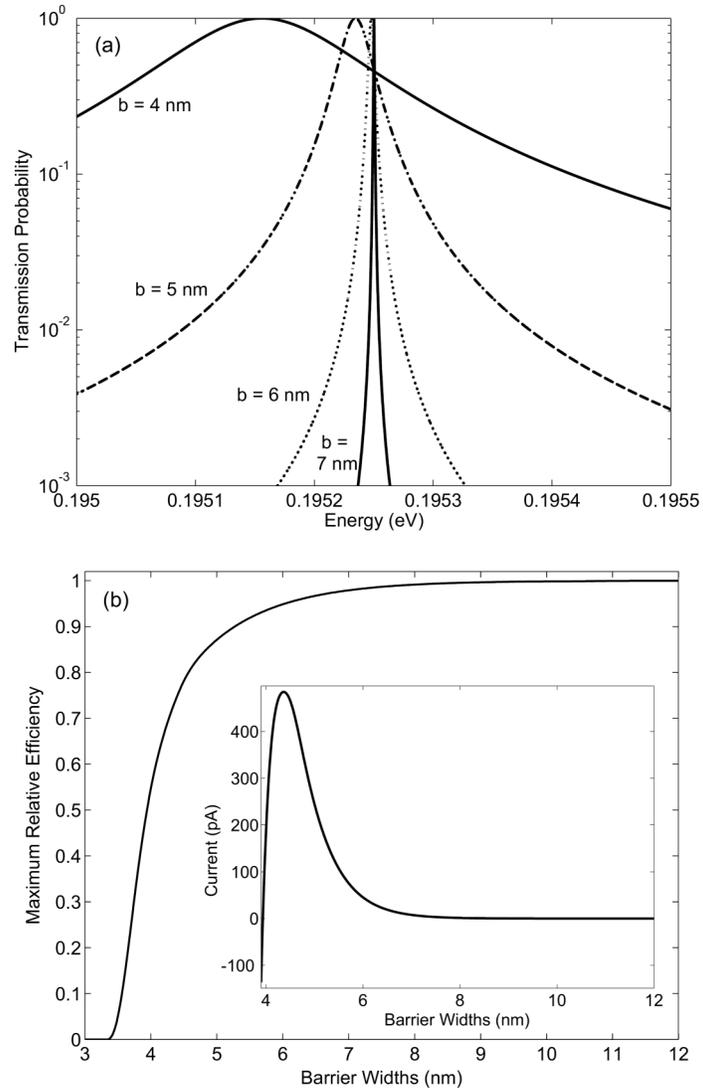

**Figure (4)** (a) The width of resonances for $w = 7$ nm and $b = 4, 5, 6$ and 7 nm (see figure 1). (b) Device efficiency relative to the Carnot value $\eta_c = (1-1K/10K) = 90\ \%$ versus $b$ for $w = 7$ nm. For each value of $b$, $\mu_C$ and $V_{bias}$ have been tuned to find the maximum efficiency value. (Inset) Device current (sampled at -1 mV) versus $b$.

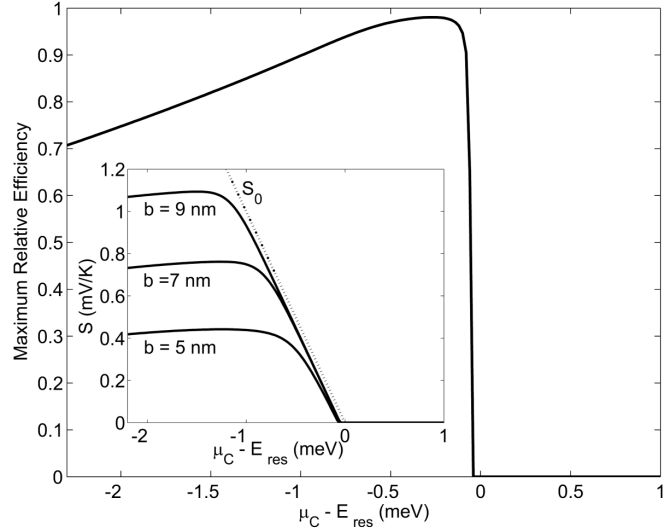

**Figure 5.** $\eta/\eta_c$ as a function of, $\mu_C$ with fixed $E_{res}$. (Inset) The Seebeck coefficient, $S$, as a function of $\mu_C$ with fixed $E_{res}$ for different barrier widths as labeled. The maximum possible Seebeck coefficient, which is associated with reversibility, $S_0$, is also included.

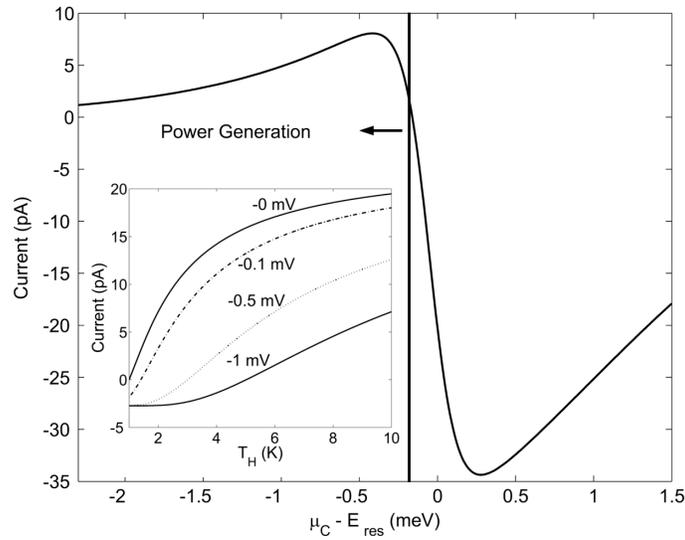

**Figure 6.** The device current sampled at $V_{bias} = -1$ mV versus the difference between the fixed resonance energy and varied electrochemical potential. (Inset) Current, sampled at $V_{bias} = 0$ mV, -0.1 mV, -0.5 mV and -1 mV as labeled, versus the hot reservoir temperature. $\mu_C - E_{res} = -0.3$ meV, which corresponds to the electrochemical potential for maximum efficiency as shown in figure 5.

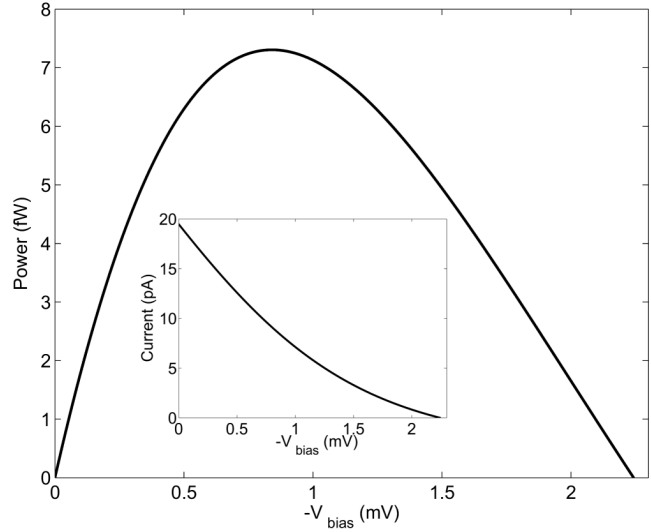

**Figure 7.** Power versus bias voltage. (Inset) Current versus the bias voltage. $\mu_C - E_{res} = -0.3$ meV.

It is very difficult to directly demonstrate power generation near Carnot efficiency in an experiment, because both $P_{out}$ and $\dot{Q}_H$ go to zero as $\eta$ approaches $\eta_c$. However, we recently showed that Carnot efficient thermoelectric power generation corresponds to a Seebeck coefficient, $S = V_{OC}/(T_H - T_C)$, that approaches a theoretically maximum value [4]

$$S_0 = \frac{1}{e}\frac{(E_{res} - \mu_H)}{T_H}. \qquad (4)$$

Here, $V_{OC}$ is the open-circuit voltage which can be measured also in the limit of negligible power. In the inset to figure 5 we show that the Seebeck coefficient approaches $S_0$ for $\eta/\eta_c \to 1$.

**5. Conclusions**

We have presented a concept study for an experimental demonstration of thermoelecric power generation near the Carnot limit. We considered a double-barrier resonant tunneling structure embedded into a InAs/InP nanowire heterostructure which can be grown using chemical beam epitaxy and seeding by size-selected gold particles [5,9]. Our results presented here show that in order to achieve efficiency approaching the Carnot value in the temperature range of 1 to 10K (where electron-phonon coupling can be suppressed), the device must restrict electrons to a narrow total energy range of less than about $kT_C \approx 0.1$ meV, which is achieved via quantization of transverse energy by the wire and longitudinal energy via resonant transmission through a double barrier structure of width $w \approx 7$ nm and with symmetric barriers of width $b \approx 7$ nm. With this device, currents of the order of 10 pA against a bias of -1 mV are predicted. We expect efficiencies approaching the Carnot value, which may be measured directly or indirectly by demonstrating thermopowers approaching the theoretical limit $S_0$.

**Acknowledgements**

This project is supported by the Office of Naval Research (ONR) and ONR Global. M.O'D. is supported by the Australian Research Council. T.E.H. is supported by a Marie Curie Incoming International fellowship from the European Commission.

**Appendix**

In the analysis above we assume that the resonant energy levels, transverse energy levels and levels due to Coulomb blockade effects are all separated from the first energy level, $\varepsilon_1$, by a few $k_B T$ so that it is accurate to assume that transport is dominated by this band only. The value of $\varepsilon_1$ is unimportant here as it simply an offset which can be accounted for by adjusting the electrochemical potential in all equations to follow. Total electron energy is therefore given by

$$E = \varepsilon_1 + E_x = \varepsilon_1 + \frac{\hbar^2 k_x^2}{2m^*} \qquad (A.1)$$

where $E_x$ and $k_x$ are the energy and wavenumber in the direction of transport respectively and $m^*$ is the effective mass (the nanowire radius is sufficient for the effective mass approximation to be used).

Under these assumptions the electric current may be calculated using the Landauer formalism as

$$I = \frac{e}{h} \int (f_H - f_C) \xi(E_x) dE_x \qquad (A.2)$$

where $e = -1.602 \times 10^{-19}$ C is the charge of an electron, $\zeta$ is the transmission probability and

$$f_{H/C} = \left(1 + \exp\left(\frac{E - \mu_{H/C}}{k_B T_{H/C}}\right)\right)^{-1} \qquad (A.3)$$

is the Fermi-Dirac distribution for electrons in the hot/cold reservoir. Heat current is calculated by noting that an electron leaving the hot/cold reservoir removes heat equal to the difference between its total energy and the electrochemical potential of the reservoir [3] giving the net heat current leaving the hot/cold reservoir as

$$\dot{Q}_{H/C} = \frac{1}{h} \int (E - \mu_{H/C})(f_H - f_C) \xi(E_x) dE_x . \qquad (A.4)$$

As was previously discussed we shall assume zero thermal conductivity from the hot to cold reservoir. The chemical potential may be approximated by the Fermi energy provided it is much larger than $k_B T$, which will be the case for this system. The efficiency of the nanowire device, acting as a heat engine is calculated as $\eta = P_{out} / \dot{Q}_H$.